\documentclass[conference]{IEEEtran}
\IEEEoverridecommandlockouts
\usepackage{cite}
\usepackage{amsmath,amssymb,amsfonts}
\usepackage{algorithmic}
\usepackage{graphicx}
\usepackage{textcomp}
\usepackage{xcolor}
\usepackage{multirow}
\usepackage{url}
\usepackage{float}
\usepackage{epstopdf}
\usepackage{CJKutf8}
\def\BibTeX{{\rm B\kern-.05em{\sc i\kern-.025em b}\kern-.08em
    T\kern-.1667em\lower.7ex\hbox{E}\kern-.125emX}}
\begin{document}

\title{Noise Classification Aided Attention-Based Neural Network for Monaural Speech Enhancement
\thanks{This work was supported by National Key R$\&$D Program of China, under Grant No. 2020AAA0104500. The corresponding author is Lu Ma. Email: malu6@tal.com, iamroad@163.com}
}


\author{\IEEEauthorblockN{Lu Ma, Song Yang, Yaguang Gong, Zhongqin Wu}
\IEEEauthorblockA{TAL Education Group \\
Beijing, China \\
\{malu6,yangsong1,gongyaguang,wuzhongqin\}@tal.com}}

\maketitle

\begin{abstract}
This paper proposes an noise type classification aided attention-based neural network approach for monaural speech enhancement. The network is constructed based on a previous work by introducing a noise classification subnetwork into the structure and taking the classification embedding into the attention mechanism for guiding the network to make better feature extraction. Specifically, to make the network an end-to-end way, an audio encoder and decoder constructed by temporal convolution is used to make transformation between waveform and spectrogram. Additionally, our model is composed of two long short term memory (LSTM) based encoders, two attention mechanism, a noise classifier and a speech mask generator. Experiments show that, compared with OM-LSA and the previous work, the proposed noise classification aided attention-based approach can achieve better performance in terms of speech quality (PESQ). More promisingly, our approach has better generalization ability to unseen noise conditions.
\end{abstract}


\section{Introduction}
Speech enhancement, usually called speech denoisng, is a task of improving speech quality and intelligibility \cite{ref_se}. It plays a key role in speech, audio and acoustic applications, such as telecom, hands-free telephone, mobile communication, etc. The feeling of voice interaction will degrade severely when noise exists, especially for complicated noise, such as babble noise and factory noise. The influence could be improved by multi-channel processing technologies if multiple microphones were available \cite{ref_micarray,ref_micarray1}. In this paper, we focus on the problem of single-channel enhancement where only one microphone is used for audio recording.

Over the past several decades, lost of methods have been proposed to handle this problem. In general, two categories of methods can be classified, namely traditional signal processing approaches and deep-learning approaches. The traditional methods, such as, spectral subtraction (SS) \cite{ref_ss}, Wiener filtering (WF) \cite{ref_wf}\cite{ref_wf1} and adaptive filtering (AF) \cite{ref_af}, are based on a predefined noise or speech statistical assumption, such as Gaussian distribution or Laplace distribution \cite{ref_gaussian}\cite{ref_distribution}\cite{ref_distribution1}. They are less effective in low Signal-to-Noise Ratio (SNR) and non-stationary noise conditions. It becomes more severe when the noise distribution is different with or deviated from the pre-assumption.

Recently, deep learning has been proved to be more effective to complex problems that
were previously unattainable with signal processing techniques. It is a data-driven supervised learning approach by learning a mapping function via observing a large number of representative pairs of noisy and noise-free speech samples. Since no
statistical assumption is made in advance, this makes it popular to bring deep neural network (DNN) methods into speech enhancement. Mostly,  time frequency (T-F) mask is used as the network learning objective \cite{ref_mask}\cite{ref_mask1}. Therefore, the estimated clean speech is obtained by multiplying this mask with the noisy spectrogram and transforming to audio waveform by inverse-short-time fourier transform (iSTFT). Various methods has adopted this kind of mask-estimation structure, such as \cite{ref_cnn,ref_cnn1,ref_lstm,ref_lstm1,ref_lstm2,ref_lstm3}. Nowadays, approaches by directly feeding raw waveforms into a neural network for enhancement and directly output audio waveform has arised, such as SEGAN \cite{ref_end2end}, WaveNet \cite{ref_end2end1}.

In this paper, we further explore the attention-based neural network structures for speech enhancement based on the previous work proposed in \cite{ref_lstm3}. Additionally, we proposed to introduce a noise type classification subnetwork into the model. It works parallelly with the denosing subnetwork. This idea is inspired by \cite{ref_class} where a subnetwork of voice activity detection (VAD) is embedded for guiding the denosing subnetwork. They share the same audio encoder and spectrogram encoder that are used for extracting high-level representation of the input audio. Then, the noise type is estimated using a LSTM encoder with attention mechanism. The generated noise context is fed to the denosing subnetwork which is constructed by another LSTM encoder with attention mechanism. The noise context is concatenated with the LSTM embedding of denoising subnetwork for attention mechanism in denoising. This way of using attention is inspired by \cite{ref_attention} where aspect embedding is concatenated with LSTM hidden embedding and used in attention for sentiment classification. In the work, causal local attention where the current frame and the previous frames within a window is considered for attention, is used for considering real-time processing scenarios. Since the noise type information is embedded into the attention mechanism for denoising, a more precise estimation could be gained. We conducted comparison experiments as the same as indicated in \cite{ref_attention}. Experiments show that, the proposed structure can consistently achieve better PESQ performance and generalization ability.

\begin{figure*}[htb]
\begin{center}
\includegraphics[width=150mm]{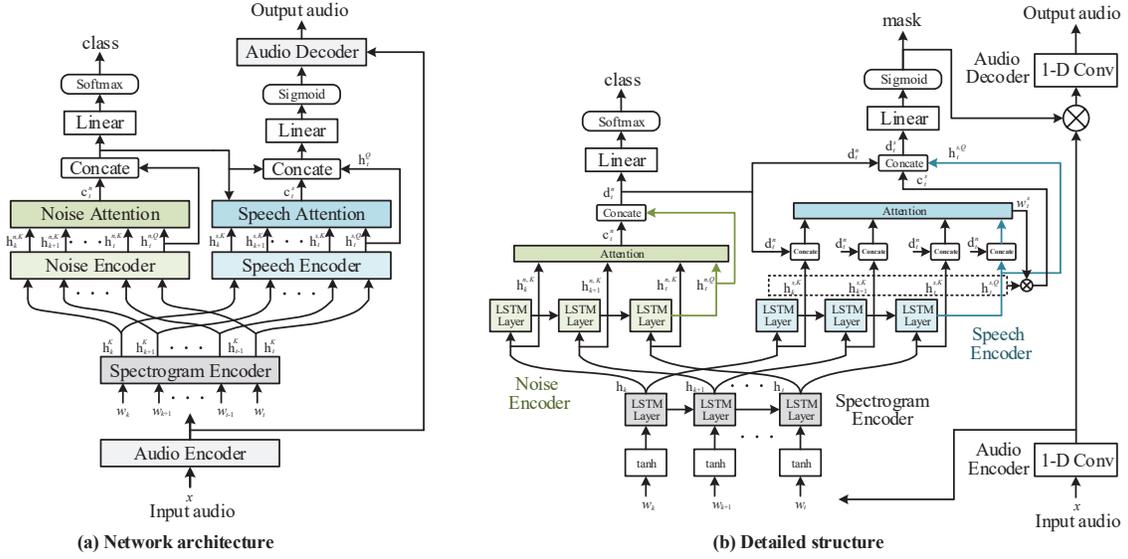}
\end{center}
\caption{Network structure of of the proposed noise classification aided attention-based model
for speech enhancement.}
\vspace*{-3pt}
\label{fig:structure}
\end{figure*}

\section{Network Structure}
The network architecture is shown in Fig. \ref{fig:structure}(a). It is constituted by an audio encoder and decoder, and a spectrogram encoder, a noise encoder with attention, a speech encoder with attention, a mask generator. The detailed model structure is shown in Fig. \ref{fig:structure}(b). Encoder module is used to transform short segments of the input waveform into their corresponding spectrograms. The spectrogram is encoded by spectrogram encoder for obtaining a high-level feature representation. Then, this feature representation is fed to two parallel branches, one for noise type classification and the other for speech enhancement. The speech waveform is then reconstructed by transforming the masked representation using a decoder.

\subsection{Audio encoder}
\label{sec:encoder}
The input audio is divided into overlapping segments of length $L$ samples. It is represented by $\mathbf{x}_{k} \in \mathbb{R}^{1 \times L}$, where $k=1, \ldots, {T}$ denotes the segment index and ${T}$ denotes the total number of segments. $\mathbf{x}_{k}$ is transformed into a $N$-dimensional representation, by a 1--D convolution operation $\mathbf{U} \in \mathbb{R}^{1 \times N}$ (denoted by $1$-$D$ $Conv$). It is formulated by a matrix multiplication as

\begin{equation} \label{eq_audio_encoder}
{\mathbf{w}}=\mathcal{H}(\mathrm{\mathbf{x}} \mathbf{U})
\end{equation}
where $\mathbf{U} \in \mathbb{R}^{N \times L}$ contains $N$ vectors (encoder basis functions) with length $L$ for each, $\mathcal{H}(\cdot)$ is the rectified linear unit (ReLU) function to ensure non-negative of the representation.

\subsection{Spectrogram encoder}
\label{sec:spec_encoder}

The spectrogram encoder extracts a high-level feature representation $\mathbf{h}$ from the input spectrogram $\mathbf{w}$:

\begin{equation} \label{eq_spec_encoder}
\mathbf{h}=\operatorname{Encoder^{spec}}(\mathbf{w})
\end{equation}
where $\mathbf{h}$ is the spectrogram embedding and fed to the following encoders for different task. In our work, we adopt LSTM as the encoder that has strong sequential modeling ability leading to superior performances in speech enhancement.

\subsection{Noise classification}
\label{sec:classification}
The noise classification subnetwork is composed of an noise feature extraction constructed by a LSTM layer and an attention mechanism and a classification module constructed by a linear layer with softmax activation function.

$\textbf{Noise\ Encoder}$ The spectrogram embedding $\mathbf{h}$ is transformed to noise embedding $\mathbf{h}^{n}$ by a LSTM layer as

\begin{equation} \label{eq_noise_encoder}
\mathbf{h}^{n}=\operatorname{Encoder^{noise}}(\mathbf{h})
\end{equation}

$\textbf{Noise\ Attention}$ The spectrogram embedding $\mathbf{h}$ and the noise embedding $\mathbf{h}^{n}$ are used for noise attention, where the noise embedding $\mathbf{h}^{n}$ acts as query, and the spectrogram embedding $\mathbf{h}$ acts as key and value. It is expressed as

\begin{equation} \label{eq_noise_attention}
\mathbf{c}^{n}=\text{Attention}\left(\mathbf{h}, \mathbf{h}^{n}\right)
\end{equation}
where $\mathbf{c}^{n}$ is the generated context vector of the noise attention.

As is shown in Fig. \ref{fig:structure}(b), a casual local attention is used to avoid any latency for speech enhancement in practice. Therefore, if we denoise a frame, $\mathbf{x}_{k}$, we calculate attention weights within a window of length $w$ using $\left[\mathbf{x}_{t-w}, \cdots, \mathbf{x}_{t}\right]$. This means that, the corresponding spectrogram embedding $\left[\mathbf{h}_{t-w}, \cdots, \mathbf{h}_{t}\right]$ and is used as the key and vale, while the noise embedding of the $t$-th frame, $\mathbf{h}_{t}^{n}$, is used as the query. Therefore, eq. (\ref{eq_noise_attention}) is rewritten as,
\begin{equation} \label{eq_noise_attention}
\mathbf{c}_{t}^{n}=\text{Attention}\left(\left[\mathbf{h}_{t-w}, \cdots, \mathbf{h}_{t}\right], \mathbf{h}_{t}^{n}\right)
\end{equation}
where, the upper subscripts of $K$ and $Q$, representing key and vale, is omitted for simplicity. Thus, a normalized attention weight $\alpha_{t,k}^{n}$ is learned:

\begin{equation} \label{eq_noise_attention1}
\alpha_{t,k}^{n}=\frac{\exp \left(\operatorname{score}\left(\mathbf{h}_{k}, \mathbf{h}_{t}^{n}\right)\right)}{\sum_{k=t-w}^{t} \exp \left(\operatorname{score}\left(\mathbf{h}_{k}, \mathbf{h}_{t}^{n}\right)\right)}
\end{equation}
We follow the correlation calculation in [27], so score $\left(\mathbf{h}_{k}, \mathbf{h}_{t}^{n}\right)=\mathbf{h}_{k}^{\top} \mathbf{W h}_{t}^{n}$. Finally, we compute the context vector as the weighted average of $\mathbf{h}_{k}$ as:

\begin{equation} \label{eq_noise_attention2}
\mathbf{c}_{t}^{n}=\sum_{k=t-w}^{t} \alpha_{t,k}^{n} \mathbf{h}_{k}
\end{equation}

$\textbf{Classification}$ The context vector of the noise attention, $\mathbf{c}_{t}^{n}$, is concatenated with the noise embedding, $\mathbf{h}_{t}^{n}$, i.e., $\mathbf{d}_{t}^{n}=\left[\mathbf{c}_{t}^{n} ; \mathbf{h}_{t}^{n}\right]$, and fed to linear layer for noise type classification,

\begin{equation} \label{eq_noise_linear}
\text{class}=\text{softmax} \left(\mathbf{W}_{e}^{n}\mathbf{d}_{t}^{n}]+\mathbf{b}_{e}^{n}\right)
\end{equation}
where the $[\cdot ; \cdot]$ denotes the concatenation of two vectors. Finally, the noise type if

\subsection{Speech denoising}
\label{sec:denoising}

The denoising subnetwork is composed of an speech feature extraction constructed by a LSTM layer and an attention mechanism and a mask generator constructed by a linear layer with sigmoid activation function.

$\textbf{Speech\ Encoder}$ The spectrogram embedding $\mathbf{h}$ is transformed to speech embedding $\mathbf{h}^{s}$ by a LSTM layer as

\begin{equation} \label{eq_speech_encoder}
\mathbf{h}^{s}=\operatorname{Encoder^{speech}}(\mathbf{h})
\end{equation}

$\textbf{Speech\ Attention}$ As is shown in Fig. \ref{fig:structure}(b), the speech embedding and the spectrogram embedding are respectively concatenated with the noise embedding, $\mathbf{d}_{t}^{n}$, i.e., $\mathbf{f}_{t}^{s}=\left[\mathbf{d}_{t}^{n} ; \mathbf{h}_{t}^{s}\right]$, and
$\mathbf{f}_{k}=\left[\mathbf{d}_{t}^{n} ; \mathbf{h}_{k}\right]$, where $k =t-w,\cdots,t$. The concatenated speech embedding $\mathbf{f}^{s}_{t}$ acts as query, and the concatenated spectrogram embedding $\mathbf{f}_{k}$ acts as key and value, and speech attention is performed by,

\begin{equation} \label{eq_speech_attention}
\mathbf{c}_{t}^{s}=\text{Attention}\left(\left[\mathbf{f}_{t-w}, \cdots, \mathbf{f}_{t}\right], \mathbf{f}_{t}^{s}\right)
\end{equation}
where $\mathbf{c}^{s}_{t}$ is the context vector of the speech attention. Again, he upper subscripts of $K$ and $Q$ is omitted for simplicity. Thus, a normalized attention weight $\alpha_{t,k}^{s}$ is learned:

\begin{equation} \label{eq_speech_attention1}
\alpha_{t,k}^{s}=\frac{\exp \left(\operatorname{score}\left(\mathbf{f}_{k}, \mathbf{f}_{t}^{s}\right)\right)}{\sum_{k=t-w}^{t} \exp \left(\operatorname{score}\left(\mathbf{f}_{k}, \mathbf{f}_{t}^{s}\right)\right)}
\end{equation}
Finally, we compute the context vector of speech attention by multiplying the spectrogram embedding, $\mathbf{h}_{k}$, with the corresponding attention weights, $\alpha_{t,k}^{s}$, by,

\begin{equation} \label{eq_speech_attention2}
\mathbf{c}_{t}^{s}=\sum_{k=t-w}^{t} \alpha_{t,k}^{s} \mathbf{h}_{k}
\end{equation}
Therefore, the noise information is embedded into the denosing subnetwork, guiding it to gain higher performance.

$\textbf{Masking}$ The context vector of the speech attention, $\mathbf{c}_{t}^{s}$, is concatenated with the speech embedding, $\mathbf{h}_{t}^{s}$ and the noise embedding, $\mathbf{d}_{t}^{n}$, and fed to a linear layer to obtain an enhancement vector $\mathbf{e}_{t}^{s}$ as

\begin{equation} \label{eq_speech_linear1}
\mathbf{e}_{t}^{s}=\tanh \left(\mathbf{W}_{e}^{s}\left[\mathbf{c}_{t}^{s} ; \mathbf{h}_{t}^{s}; \mathbf{d}_{t}^{n}\right]+\mathbf{b}_{e}^{s}\right)
\end{equation}
where the $[\cdot ; \cdot ; \cdot]$ denotes the concatenation of three vectors. Finally, we form a mask of the input feature $\mathbf{x}_{t}$, and get the final enhanced speech spectrogram $\mathbf{y}_{t}$ as

\begin{equation} \label{eq_speech_linear2}
\mathbf{y}_{t}=\mathbf{w}_{t} \odot \operatorname{sigmoid}\left(\mathbf{W}_{m}^{s} \mathbf{e}_{t}^{s}+\mathbf{b}_{m}^{s}\right)
\end{equation}

\subsection{Audio decoder}
\label{sec:decoder}
The decoder reconstructs the waveform from masked spectrogram using a 1--D transposed convolution operation. It is reformulated as matrix multiplication as
\begin{equation}
 \hat{\mathbf{x}_{t}}=\mathbf{y}_{t} \mathbf{V}
\end{equation}
where $\hat{\mathbf{x}} \in \mathbb{R}^{1 \times L}$ is the reconstruction of $\mathbf{x}$ and the rows in $\mathbf{V} \in \mathbb{R}^{N \times L}$ are the decoder basis functions, each with length $L$ samples. The overlapping reconstructed segments are summed together
to get the final waveforms.

\subsection{Training objective}
\label{sec:loss_cun}
The training objective is minimizing the loss function which is obtained by combining the mean square error (MSE) of the estimated waveform and the cross-entropy of between the estimated classification label, formulated by,
\begin{equation} \label{eq:mask_loss}
\left\{\begin{array}{l}
\operatorname{Loss}_{total}=(1-\alpha) \cdot {\operatorname{Loss}_{\rm{MSE}}} +\alpha \cdot {\operatorname{Loss}_{\rm{CE}}} \\
{\operatorname{Loss}_{\rm{MSE}}} = \sum_{n}^{T}\left(\hat{s}(n)-s(n)\right)^{2} \\
\operatorname{Loss}_{\mathrm{CE}}=-\sum_{n}^{T} \sum_{i=1}^{C} p\left(\hat{c}_{i}(n)\right) \log \left(p\left(c_{i}(n)\right)\right)
\end{array}\right.
\end{equation}
where ${Loss_{\rm{MSE}}}$ is the MSE between the estimated waveform $\hat{s}(n)$ and the referenced one ${s}(n)$, ${Loss_{\rm{CE}}}$ is the cross-entropy between the estimated classification $\hat{c_i}(n)$ and the referrenced classification ${c_i}(n)$, $i \in [1,2,..,C]$ is the the categories, $n$ is the time index, $\alpha \in[0,1]$ is the weighting factor.

\section{Experiments}

\subsection{Datasets}
We follow the same procedure as is indicated in \cite{ref_lstm3} to create synthetic datasets. $Librispeech$ train-clean-360 is used as the clean speech dataset which has 921 speakers and the total number of 104014 audio samples. $Nonspeech\ Sounds$ dataset \cite{ref_nonspeech} is used as the training noise dataset. We use this noise dataset, because this noise dataset is already classified to 20 categories\footnote{http://web.cse.ohio-state.edu/pnl/corpus/HuNonspeech/HuCorpus.html}.
We randomly choose 500 speakers from the 921 speakers in the train-clean-360 as the training speakers, and the remaining 421 speakers are used for unseen speakers testing.
Specifically, we randomly select a clean speech file from the 500 speakers and a noise file from the nonspeech corpus, generating 21407 utterances in total. Then we randomly select an SNR between 0dB and 20dB, and mix these two files to create a noisy file according to the selected SNR. We divide the generated 21407 noisy files into 13407, 4000 and 4000 for training, validation and testing. Neural network training is conducted using the training set and the loss on the validation set is examined as the convergence condition. The original testing set is named as Test-0. We further create 4 new test sets (Test-1,2,3,4) each with 4000 utterances randomly chosen from the remaining 431 speakers. In Test-1,2, noise files are from the noise pool, Musan, as Test-0. These Test-3,4 are generated by adding different noises (with the training set) from the $Musan$ corpus \cite{ref_musan}. Since the CHIME3 dataset used in \cite{ref_lstm3} is not open access, we use the Musan corpus instead. The SNR of these test sets are illustrated as Table \ref{tab:dataset}.

\begin{table}[htb]
\setlength\tabcolsep{3pt}
\setlength{\abovecaptionskip}{0.2cm}
\setlength{\belowcaptionskip}{0.2cm}
\centering
\caption{Conditions of noise and SNR for datasets. Utterances in
Train, Valid and Test-0 are from the same multi-speaker set; while
utterances in Test-1,2,3 and 4 from another set of speakers.}
\label{tab:dataset}
\begin{tabular}{c|c|c|c|c|c|c|c}
\hline
\text{Set} & \text{Train} & \text{Valid} & \text{Test-0} & \text{Test-1} & \text{Test-2} & \text{Test-3} & \text{Test-4} \\ \hline \hline
\multirow{2}{*}{Noise} & Non- & Non- & Non- & Non- & Non- & \multirow{2}{*}{Musan} & \multirow{2}{*}{Musan} \\
& speech & speech & speech & speech & speech & & \\ \hline
Speakers & 500 & 500 & 500 & 431 & 431 & 431 & 431 \\ \hline
SNR & 0-20dB & 0-20dB & 0-20dB & 0-20dB & -5-0dB & 0-20dB & -5-0dB \\ \hline
\end{tabular}
\end{table}

\subsection{Configurations}
In our experiments, waveforms at 16 kHz sample rate were directly served as the inputs. These models are initialized with the normalized initialization. The loss function used for training the network is Eq. (\ref{eq:mask_loss}). $\emph{Adm}$ algorithm was used for training with an exponential learning rate decaying strategy, where the learning rate starts at $1$$\times$$10^{-4}$ and ends at $1$$\times$$10^{-8}$. The total number of epochs was set to be 200. The criteria for early stopping is no decrease in the loss function on validation set for 10 epochs.

We compare our approach (namely CA-Att-LSTM) with conventional OM-LSA method, an LSTM approach without attention mechanism (namely Pure-LSTM) and the attention-based LSTM model in \cite{ref_lstm3} (namely Att-LSTM). The Pure-LSTM has two layers whose first layer size is $H$ as listed in \ref{tab:configuration} and the second layer has 128/256/512 cells. As for our CA-Att-LSTM model, two structures are compared according to whether the noise context $\mathbf{d}_{t}^{n}$ is concatenated with the speech embedding for attention, i.e. CA-Att-LSTM without concatenation (namely CA-Att-LSTM1) and with concatenation (namely CA-Att-LSTM2). The parameters configuration of the proposed network is listed in Table \ref{tab:configuration} where $C$ denotes the number of categories of noise type.

\begin{table}[htb]
\setlength\tabcolsep{2pt}
\setlength{\abovecaptionskip}{0.2cm}
\setlength{\belowcaptionskip}{0.2cm}
\centering
\caption{Network configuration}
\label{tab:configuration}
\begin{tabular}{c|c|c}
\hline
\text{Symbol} & \text{Description} & \text{Value} \\ \hline \hline
$N$ & Number of filters in encoder and decoder, Eq. (\ref{eq_audio_encoder}) & 512 \\ \hline
$L$ & Length of the filters (in samples), Eq. (\ref{eq_audio_encoder}) & 160 \\ \hline
$H$ & Hidden size of the spectrogram encoder, Eq. (\ref{eq_spec_encoder}) & 256 \\ \hline
$H^n$ & Hidden size of the noise encoder, Eq. (\ref{eq_noise_encoder}) & 60,112,224 \\ \hline
$H^s$ & Hidden size of the speech encoder, Eq. (\ref{eq_speech_encoder}) & 60,112,224 \\ \hline
$E^n$ & Output size of the linear layer, Eq. (\ref{eq_noise_linear}) & $C$$=$$20$ \\ \hline
$E^s$ & Output size of the linear layer, Eq. (\ref{eq_speech_linear1}) & 256 \\ \hline
$F$ & Output size of the linear layer, Eq. (\ref{eq_speech_linear2}) & 256 \\ \hline
$w$ & Window size of causal local attention, Eq. (\ref{eq_noise_attention}, \ref{eq_speech_attention}) & 5,15,30 \\ \hline
\end{tabular}
\end{table}

\subsection{Results}
We first analyze the performance of the baseline methods and the proposed methods on Test-0. The results of PESQ are summarized in Table \ref{tab:test0} where the averaged PESQ of the input noisy audio is $\textbf{1.67}$, and the PESQ after OM-LSA is $\textbf{1.75}$. We can clearly see that all the attention-based methods outperform the two baselines without attention (i.e., the OM-LSA and the Pure-LSTM methods) consistently for different size of parameters, which indicates that introducing attention mechanism to neural network based speech enhancement is beneficial. This indication is consistent with that of \cite{ref_lstm3}. Moreover, by introducing a noise classification subnetwork into the denoising network, our models (i.e., the CA-Att-LSTM1 and the CA-Att-LSTM2 methods) gain better performance in all configurations. This means that the noise information introduced to the network could guide the model to make better denoising. Additionally, by feeding the classification embedding to the speech encoder for attention, a higher PESQ gain can be obtained. This reveals that the noise classification can guide the speech attention to make better estimation.

In the experiments, to view the influence of window size of the causal local attention to the denoising performance, the window size was set to 5, 15 and 30 for comparisons. As is shown in the table, the best performance is achieved when $w=5$ and larger $w$ gains no further improvement. This is also consistent with that of \cite{ref_lstm3}. Here, the window size of the causal local attention used for the noise encoder and the speech encoder is same. Specifically, it can be configured with different window size. We left it for the future research.

\begin{table}[htb]
\centering
\caption{PESQ of different models on Test-0.}
\label{tab:test0}
\begin{tabular}{c|c|c|c|c|c}
\hline
\multirow{2}{*}{$w$} & \multirow{2}{*}{\text{lstm-size}} & \text{Pure-}  & \text{Att-} & \text{CA-Att-} & \text{CA-Att-}        \\
 & & \text{LSTM}  & \text{LSTM} & \text{LSTM1} & \text{LSTM2}        \\ \hline \hline
\multirow{3}{*}{5} & 128/112/60/60    & 2.34   & 2.41   & 2.43  & 2.56   \\ \cline{2-6}
                   & 256/224/112/112   & 2.43   & 2.50   & 2.52  & 2.63 \\ \cline{2-6}
                   & 512/448/224/224   & 2.50   & 2.59   & 2.64  & 2.72 \\ \cline{1-6}
\multirow{3}{*}{10} & 128/112/60/60   & 2.34   & 2.39   & 2.40  & 2.55  \\ \cline{2-6}
                   & 256/224/112/112   & 2.43   & 2.51   & 2.47  & 2.66 \\ \cline{2-6}
                   & 512/448/224/224   & 2.50   & 2.57   & 2.63  & 2.71  \\ \cline{1-6}
\multirow{3}{*}{15} & 128/112/60/60   & 2.34   & 2.42   & 2.46  & 2.52  \\ \cline{2-6}
                   & 256/224/112/112   & 2.43   & 2.49   & 2.58  & 2.64 \\ \cline{2-6}
                   & 512/448/224/224   & 2.50   & 2.57   & 2.61  & 2.71 \\ \hline
\end{tabular}
\end{table}

\begin{figure}[H]
\centering
\includegraphics[width=75mm,height=225mm]{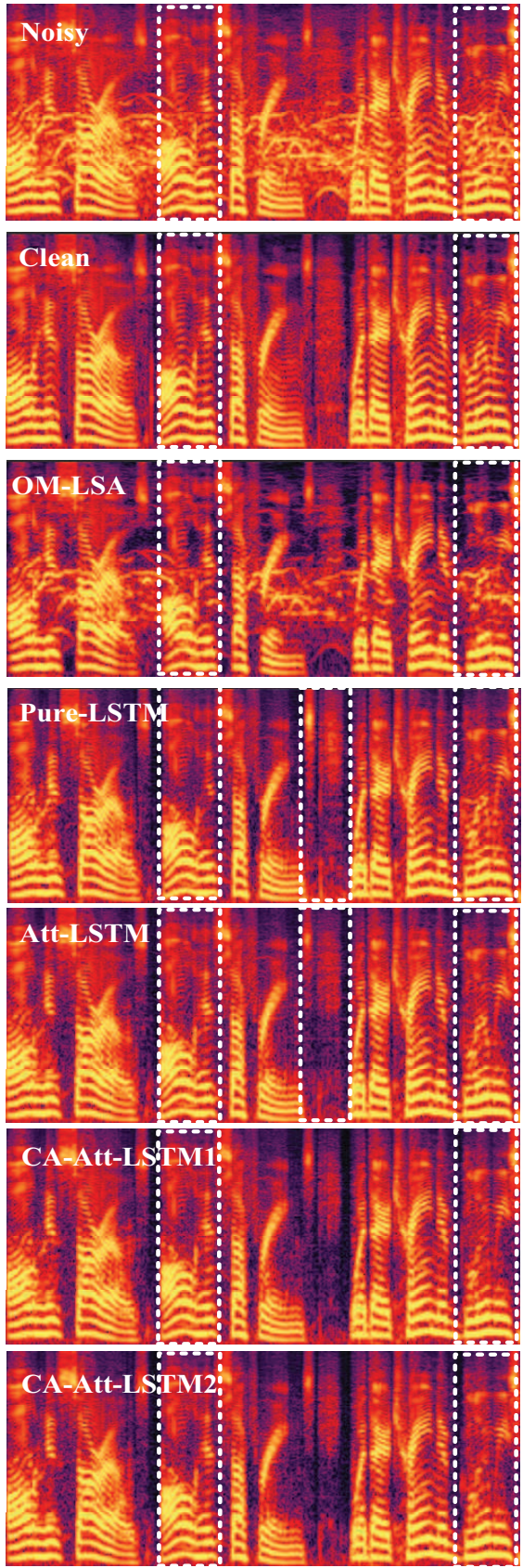}
\caption{Spectrograms comparisons for different methods.}
\label{fig:waves}
\end{figure}

To further showcase the effects of speech enhancement for different methods, a speech utterance (spectrum) randomly selected from Test-0 is shown in Fig \ref{fig:waves}. The clean speech is contaminated with `crowd' noise. As is shown in the figure and indicated by \cite{ref_lstm3}, the traditional OM-LSA method cannot handle this kind of non-stationary noise properly, and the LSTM approach can significantly reduce noise but still with some noise residuals. On the contrast, all the attention-based methods can remove the noise properly. Moreover, since noise type can be estimated and guided to denoise, our model gives a pretty good reconstruction. Some of our testing examples can be found from our repository$\footnote{\url{https://github.com/ROAD2018/noise_aware_attention_based_denoiser}}$.

To validate the generalization capability of different approaches, experimental results on Test-0,1,2,3,4 are summarized in Table \ref{tab:test1_4} for comparisons. The window size of the causal local attention used in the models are $w = 5$. The LSTM cell size is 512, 448, 256 for the pure LSTM model, the attention-based LSTM model and our models, respectively. The PESQ of the raw noisy audio are $\textbf{1.67}$, $\textbf{1.67}$, $\textbf{1.26}$, $\textbf{1.84}$, $\textbf{1.21}$ for Test-0,1,2,3,4, respectively.
We notice that since the dataset mismatch between the training and the testing, all the models have performance degradation when tested on Test-1,2,3,4 compared with Test-0. It is worth noting that our model (i.e., CA-Att-LSTM2) gain the same PESQ vale between Test-0 and Test-1. This may indicate that our CA-Att-LSTM2 model is more robust to unseen speakers.

Moreover, the models trained on the data with 0$\sim$20dB SNR have significant performance degradation on the test set with -5$\sim$0dB SNR (Test-2). Further with both mismatched noises and SNR conditions, large performance degradation can be observed on Test-4. But among all the methods, attention-based models have shown better generalization ability and our noise classification aided models gain better performance.

\begin{table}[htb]
\centering
\caption{PESQ of different models on Test-0,1,2,3,4.}
\label{tab:test1_4}
\begin{tabular}{c|c|c|c|c|c}
\hline
\multirow{2}{*}{\text{Set}}  & \text{OM-} & \text{Pure-}  & \text{Att-} & \text{CA-Att-} & \text{CA-Att-}  \\
& \text{LSA} & \text{LSTM}  & \text{LSTM} & \text{LSTM1} & \text{LSTM2}  \\ \hline \hline
\text{Test-0}   & 1.75    & 2.50    & 2.59  & 2.68    & 2.72   \\ \hline
\text{Test-1}   & 1.76    & 2.46    & 2.57  & 2.66    & 2.72     \\ \hline
\text{Test-2}   & 1.29    & 1.82    & 1.84  & 1.86    & 1.89      \\ \hline
\text{Test-3}   & 1.90    & 2.05    & 2.11  & 2.12    & 2.15      \\ \hline
\text{Test-4}   & 1.23    & 1.31    & 1.34  & 1.38    & 1.41      \\ \hline
\end{tabular}
\end{table}

\section{Discussions}
Through the above experiments, we can conclude that the noise classification can indeed assist to denoise. Therefore, to make better denoise, a possible way is to collect as many noise type as possible for training. In this paper work, we directly used the categories classified in advance. While, in a more realistic scenario, the noise should be classified automatically. This can be realized by clustering the embedding output from a pre-trained audio encoder, such as Speech-VGG \cite{ref_speechvgg}.

\section{Conclusions}
An noise classification subnetwork is introduced into the attention-based neural network for speech enhancement based on the previous work. The embedding from the classification is used together with the denoising embedding for causal local attention, therefore, guiding the network to maker better denoising. The performance of the proposed network is validated and compared with the previous work and OM-LSA and an pure LSTM approaches, obtaining higher PESQ gain. Moreover, the generalization ability to unseen noise conditions is also validated. In the
future, we will explore other encoder structure other than LSTM for speech enhancement, and bring the attention mechanism into multi-channel speech enhancement, and further integrate it with speech recognition.


\vspace{12pt}
\color{red}

\end{document}